\documentclass[
aps,%
12pt,%
final,%
notitlepage,%
oneside,%
onecolumn,%
nobibnotes,%
nofootinbib,%
superscriptaddress,%
noshowpacs]%
{revtex4}
\usepackage{graphicx}
\usepackage{color}

\usepackage[mathscr]{euscript}
\usepackage{amsmath}
\usepackage{graphicx}
\usepackage{subfig}
\usepackage{dcolumn}
\usepackage{caption}
\usepackage{bm}
\usepackage{epsfig}
\usepackage{amssymb,latexsym,mathrsfs}
\usepackage{hyperref}
\usepackage{float}
\usepackage{diagbox}

\hypersetup{ linktoc=all,
    colorlinks, linkcolor={palatinateblue},
    citecolor={brightpink}, urlcolor={amaranth}
}
\usepackage[labelfont=bf,textfont=normalfont,singlelinecheck=off,justification=raggedright]{caption}
\definecolor{vividviolet}{rgb}{0.62, 0.0, 1.0}
\definecolor{amaranth}{rgb}{0.9, 0.17, 0.31}
\definecolor{palatinateblue}{rgb}{0.15, 0.23, 0.89}
\definecolor{brightpink}{rgb}{1.0, 0.0, 0.5}
\definecolor{cornflowerblue}{rgb}{0.39, 0.58, 0.93}
\definecolor{deepcarminepink}{rgb}{0.94, 0.19, 0.22}
\definecolor{radicalred}{rgb}{1.0, 0.21, 0.37}

\newcommand{\be}{\begin{equation}}
\newcommand{\ee}{\end{equation}}
\newcommand{\bs}{\begin{split}} 
\newcommand{\bea}{\begin{eqnarray}}
\newcommand{\eea}{\end{eqnarray}}

\newcommand{\ayan}[1]{\textcolor{red}{[{\bf Ayan}: #1]}} 

\renewcommand{\d}[1]{\ensuremath{\operatorname{d}\!{#1}}}
 
\begin{document}

\noindent {\it Astronomy Reports, 2021, Vol. 98, No. 1}
\bigskip\bigskip  \hrule\smallskip\hrule
\vspace{35mm}


\title{Dual-temperature acceleration radiation\footnote{Paper presented by MG at the Fourth Zeldovich 
meeting, an international conference in honor of Ya. B. Zeldovich held in Minsk, Belarus on September 7--11, 2020. Published by the 
recommendation of the special editors: S. Ya. Kilin, R. Ruffini and G. V. Vereshchagin.}}

\author{\bf \copyright $\:$  2021.
\quad \firstname{M.~R.~R.}~\surname{Good}}%
\email{michael.good@nu.edu.kz}
\affiliation{Nazarbaev University, Nur-Sultan, Qazaqstan}%

\author{\bf \firstname{A.}~\surname{Mitra}}
\email{ayan.mitra@nu.edu.kz}
\affiliation{Nazarbaev University, Nur-Sultan, Qazaqstan}%
\affiliation{Kazakh-British Technical University, Almaty, Qazaqstan }
\author{\bf \firstname{V.}~\surname{Zarikas}}
\email{vasileios.zarikas@nu.edu.kz}
\affiliation{Nazarbaev University, Nur-Sultan, Qazaqstan}%

\begin{abstract}

 We solve for a system that emits acceleration radiation at two different temperatures.  The equilibrium states occur asymptotically in Planck distributions and transition non-thermally. The model is simple enough to obtain a global solution for the radiation spectrum analytically. We present it as a potentially useful model for investigation of non-thermal vacuum acceleration radiation in the presence of final(initial) asymptotic thermodynamic horizon(less) states in equilibrium.
\end{abstract}

\maketitle

\section{Introduction}
As one of the simplest theoretical models of acceleration radiation, the moving mirror of DeWitt \cite{DeWitt:1975ys}, and Davies-Fulling \cite{Davies:1976hi,Davies:1977yv} elucidates the process of field perturbation by a perfectly reflecting accelerated boundary, which transmits both particles and energy. The mirror model has developed \cite{carlitz1987reflections} into a textbook case \cite{Birrell:1982ix, Fabbri} of vacuum acceleration radiation, notably used in analog to describe Hawking radiation from black holes \cite{good2013time}.  On the experimental side, it has been a decade since moving mirror radiation has been detected 
\cite{Wilson_2011} and further interesting observational setups are planned \cite{Dodonov}.  Recent studies are making progress on a variety of topics including accelerating boundary entropy production \cite{Chen:2017lum}, relativistic plasma detection \cite{Chen:2015bcg, a}, de Sitter cosmology correspondence \cite{Good:2020byh}, motion induction \cite{Silva:2020odg}, extremal black holes \cite{good2020extreme}, and Casimir free-fall \cite{x,y,z} and other future prospects emanating from black hole information loss paradox \cite{mitra}.

One aspect of the moving mirror model that has not been explored is transition between equilibrium radiation states.  Equilibrium in nature is more of an exception than the rule, and
spectral changes (which constitute a significant portion of interesting radiative phenomena) take place in non-equilibrium conditions. Thus there is much to be learned about the complex evolutionary emission process occurring far from equilibrium.

Moreover, while very little is understood about the general aspects of non-equilibrium radiative systems, equilibrium spectra (e.g. Bose-Einstein statistics of a thermal Planck distribution) have been much studied and have been shown to display general universal features understood dynamically \cite{carlitz1987reflections}. This universality (e.g. in black hole thermalization) is thought to macroscopically emerge from large-scale fluctuations in such a robust way that one can expect that similar mathematical machinery and physical mechanisms will work to describe non-equilibrium radiative systems as well \cite{Zoltan}. Thus, investigating the similarities and differences of equilibrium and non-equilibrium emission may help to discover the distinguishing but still robust properties of non-equilibrium radiation.  A simple soluble system with both phases can help fill this gap.  We present such a model and ask these questions:
\begin{itemize}
    \item What kind of dynamics can be responsible for radiation at different temperatures?
    \item How can we explicitly illustrate temperature with and without a horizon?
    \item What is the spectrum describing the intermediate non-thermal phase?
   \end{itemize}

Our paper is organized as follows: in Sec.\ \ref{sec:motion}, we review the details of the proposed accelerated mirror trajectory, computing only the crucial and minimally needed relativistic dynamical properties: rapidity, speed, and acceleration. In Sec.\ \ref{sec:energy}, we derive the energy flux radiated by analysis of the quantum stress tensor. In Sec.\ \ref{sec:particles}, we derive the particle spectrum, finding a unique form for the radiation. Throughout we use natural units, $\hbar = c = 1$.
 
\section{Trajectory Motion}\label{sec:motion}
\begin{figure}
        \centering
           \subfloat[]{%
         \includegraphics[width=2.9 in]{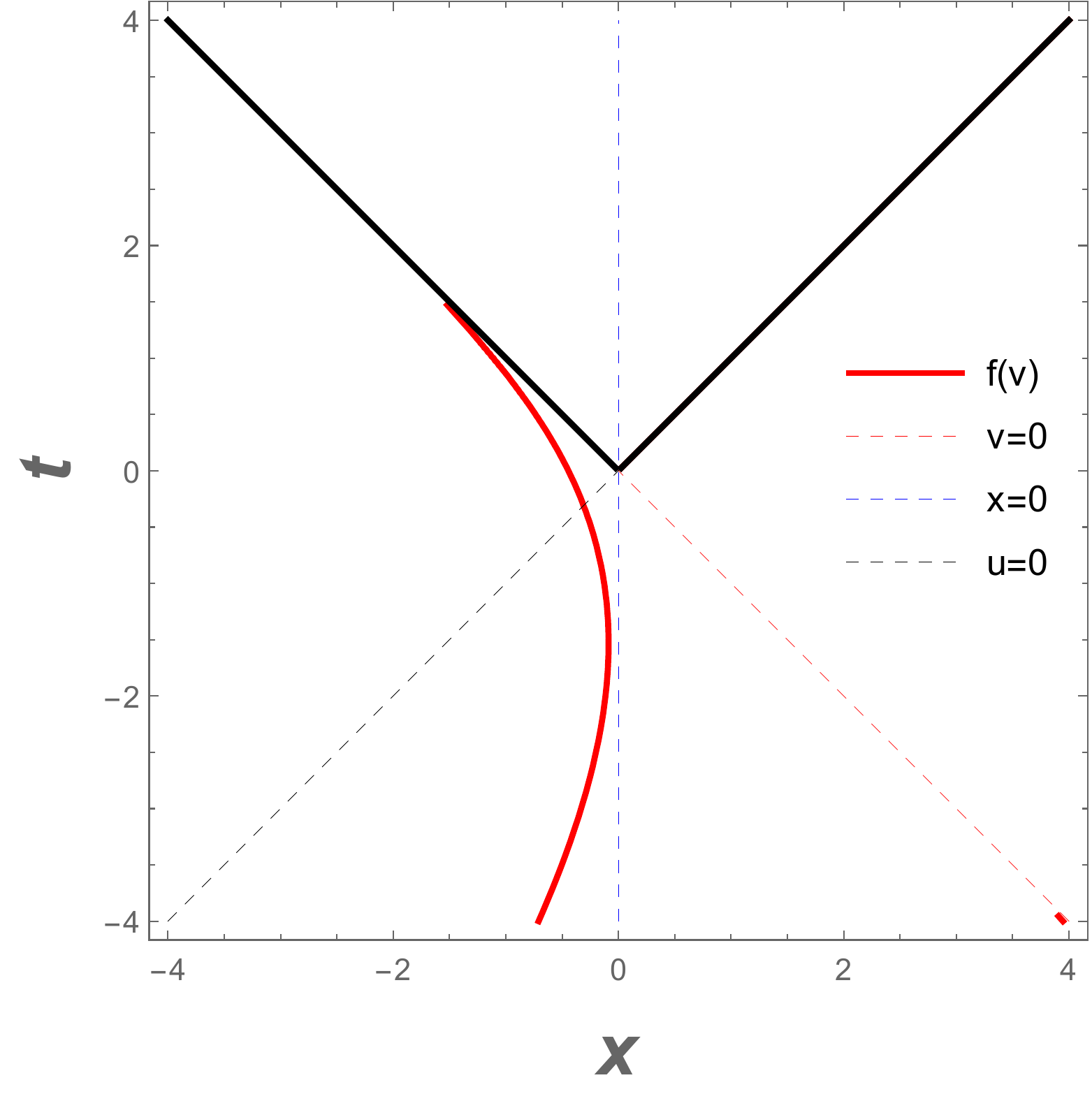}%
              \label{Fig1}%
              \qquad
           } 
           \subfloat[]{%
              \includegraphics[width=2.9 in]{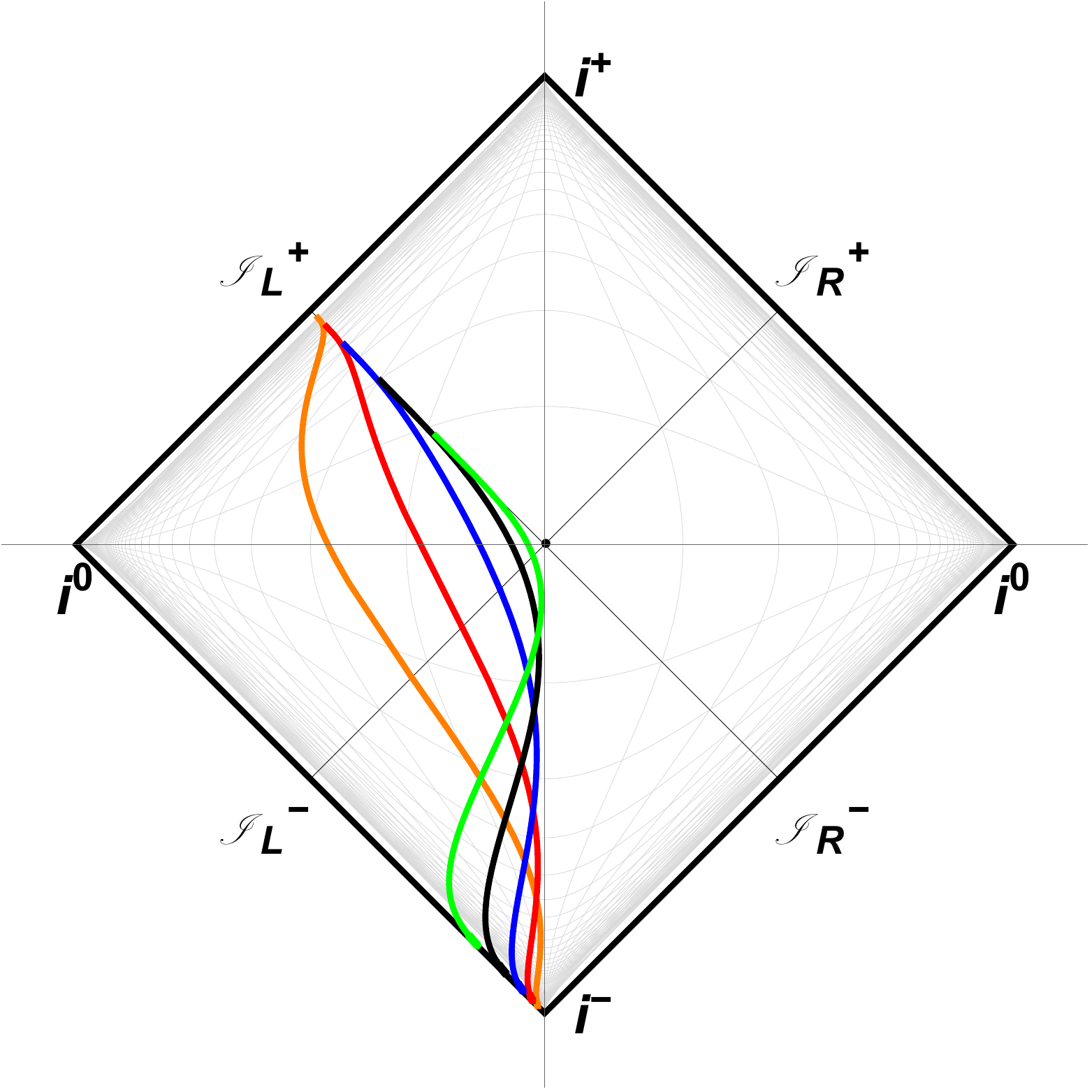}%
              \label{Fig2}%
           }
           \caption{ \textbf{Left} : A spacetime diagram of the mirror trajectory, Eq.~(\ref{f(v)}) in a contour plot.   
It starts off asymptotically inertial with zero acceleration and light-speed velocity and decelerates eventually reaching zero speed and then accelerates again approaching the speed of light receding with asymptotic infinite acceleration. This divergence happens along the null ray $v_H$ which is the advanced time horizon.  Here we have set $v_H = 0$. Notice how field modes with $v<v_H$ moving to the left will always hit the mirror.  Those field modes with $v>v_H$, never hit the mirror due to the horizon, geometrically illustrating information loss to an observer at $\mathscr{I}^+_R$.  The appearance of an initial horizon is an illusion in the spacetime diagram because the mirror starts asymptotically time-like which can be seen in the Penrose diagram of Figure \ref{Fig2}. The future light-cone centered at $t=x=0$ is marked in solid black. Here $\kappa=1$. \textbf{Right} : A Penrose diagram of the mirror trajectory, Eq.~(\ref{f(v)}). The mirror is moving at light-speed as $v\to(-\infty,0^-)$. Starting asymptotically time-like in the past, it proceeds to accelerate to an asymptotic light-like horizon at $v_H=0$.    The various colors correspond to different system scaling with $\kappa = 1/2,1,2,4,8$ from orange, red, blue, black, to green. }
           \label{F1}
    \end{figure}

We start with light-cone coordinates $(u,v)$, retarded time $u=t-x$, to express the one-parameter ($\kappa$), $1+1$-dimensional trajectory of the moving mirror as 
\be f(v) = -\frac{1}{\kappa}\ln \left[\kappa  v (\kappa  v-1)\right],\label{f(v)}\ee
where $f$ is the retarded time position.  Here $f$ is a function, not a coordinate, so the symbol ``$u$" is not used. Eq.~(\ref{f(v)}) is the retarded time `position' or trajectory of the mirror where the independent variable is advanced time $v= t+x$.  A spacetime plot with time $t$ on the vertical axis is given of the trajectory in Figure \ref{Fig1}. A conformal Penrose diagram is plotted in Figure \ref{Fig2}.  What are the important dynamics of our apriori chosen trajectory, Eq.~(\ref{f(v)})? 

\subsection{Rapidity, Speed, Acceleration}
\begin{figure}[H]
\centering
\includegraphics[width=3.1 in]{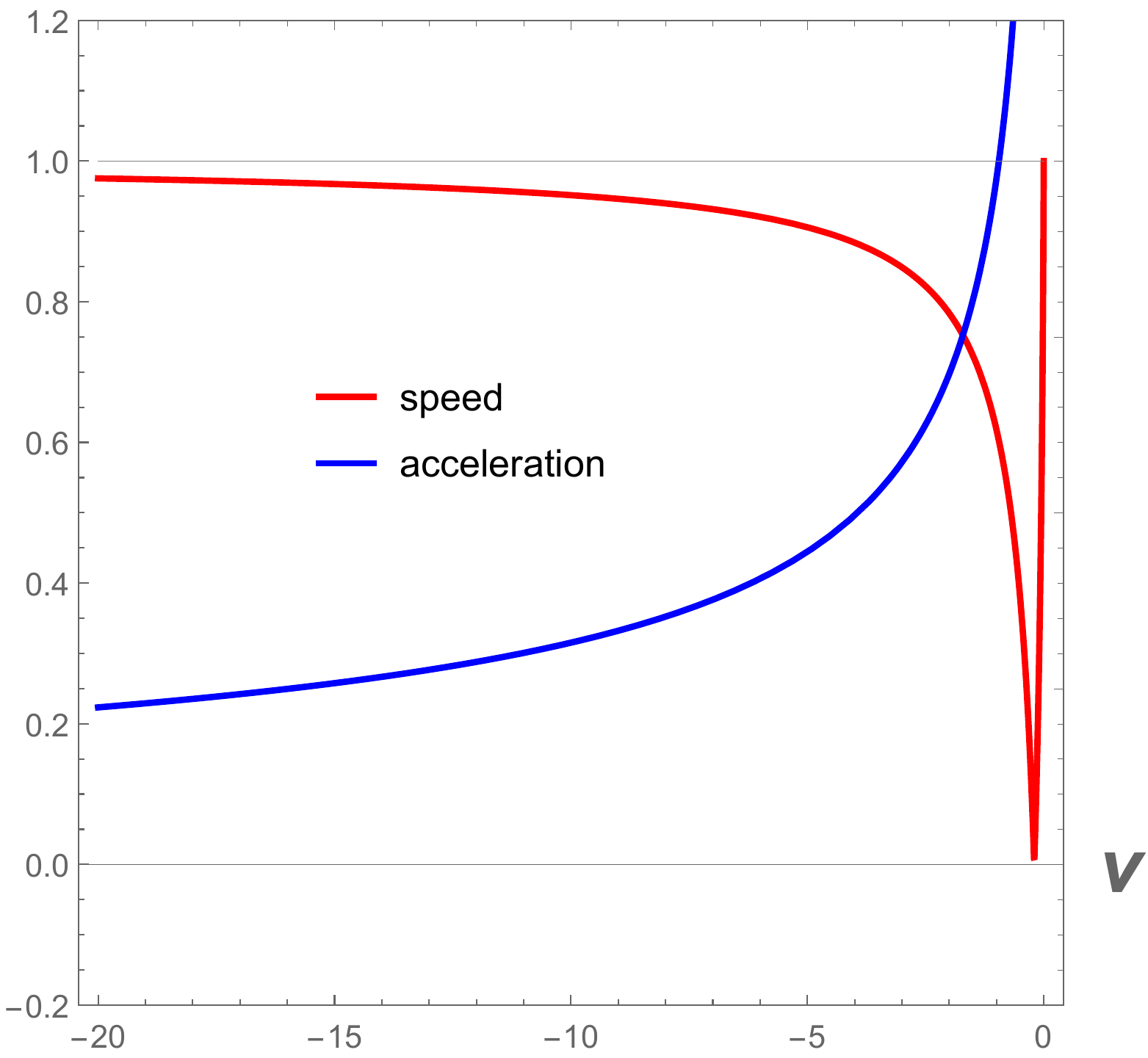}
\caption{The speed, $|V(v)|$ and proper acceleration $|\alpha(v)|$, as a function of light-cone coordinate advanced time $v=t+x$ for the mirror trajectory, Eq.~(\ref{f(v)}).  It is readily seen that at $v=0^-$, the velocity, $V$, is asymptotically the speed of light, $|V|\to c=1$ and the acceleration diverges, $|\alpha| \to \infty$. Here $\kappa=8$ for illustration. The key take-away from this graph is the asymmetry between the asymptotic states of the mirror: the past state and horizon state are both asymptotically light speed, but only the $v\to v_H = 0^-$ has infinite acceleration with a light-like horizon.}\label{Fig3}
\end{figure}

We compute the rapidity $\eta(v)$, by $2\eta(v) \equiv  -\ln f'(v)$ where the prime is a derivative with respect to the argument, plugging in Eq.~(\ref{f(v)}), 
\be \eta(v) = -\frac{1}{2} \ln \left(\frac{1-2 \kappa  v}{\kappa  v (\kappa  v-1)}\right).\label{eta(v)}\ee
With rapidity, we may easily compute the velocity, $V \equiv \tanh \eta$, plugging in Eq.~(\ref{eta(v)}),
\be V(v) = -\tanh \left(\frac{1}{2} \ln \left(\frac{1}{1-\kappa  v}-\frac{1}{\kappa  v}\right)\right), \label{V(v)}\ee
and the proper acceleration which follows from $\alpha(v)\equiv  e^{\eta(v)} \eta'(v)$, using Eq.~(\ref{eta(v)}) again, giving
the leading order terms of the acceleration at both $v\to-\infty$ and $v\to 0^-$, respectively,
\be \alpha_{\textrm{past}}(v) = -\frac{1 }{2 \sqrt{2}} \sqrt{-\frac{\kappa }{v}},\quad \alpha_{\textrm{horizon}}(v) = -\frac{1}{2} \sqrt{-\frac{\kappa }{v}}. \ee
The horizon proper acceleration has a greater magnitude by  $\alpha_{\textrm{horizon}} = \sqrt{2} \alpha_{\textrm{past}}$, which intuitively suggests a hotter radiative temperature. The magnitude of the velocity, Eq.~(\ref{V(v)}), along with the proper acceleration, $|\alpha(v)|$, 
are plotted in Figure \ref{Fig3}.

\section{Energy Flux}\label{sec:energy}
\subsection{Energy Flux}
\begin{figure}[H]
\centering
\includegraphics[width=3.5 in]{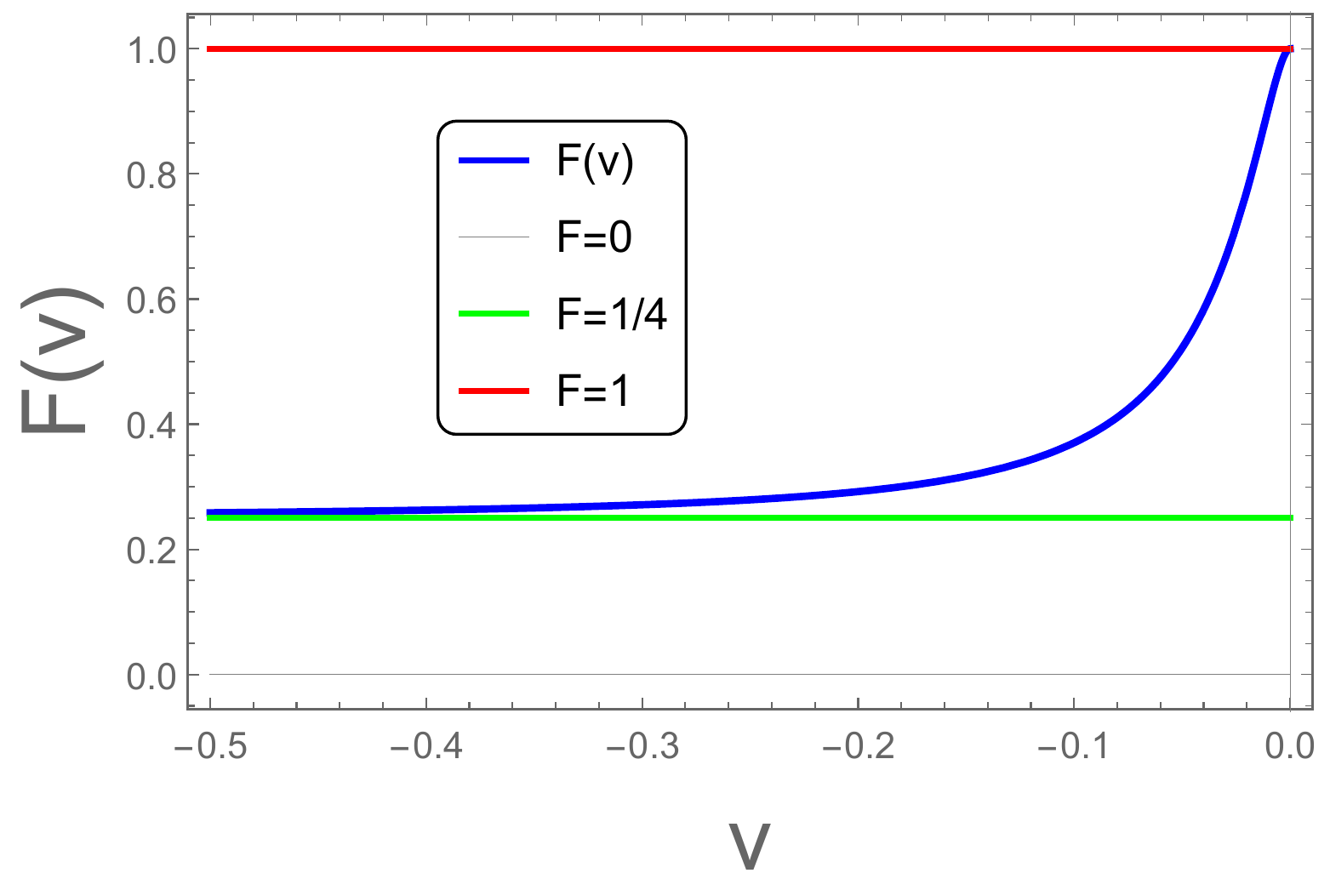}
\caption{The energy flux, Eq.~(\ref{F(v)exact}), is asymptotically constant at both $v\to (-\infty, 0^-)$. The scale has been set so that the maximum flux is $F_{\textrm{max}} = 1$, where $\kappa^2 = 48\pi$. Notice that in the asymptotic past, $v\to -\infty$, the flux is $F = \kappa^2/192\pi$, or in our scale, $F=1/4$. The radiation transitions from constant flux in the past to constant flux near the horizon (by a factor of $4$).  The temperature correspondingly increases (by a factor of $2$). }\label{Fig4}
\end{figure}

The quantum stress tensor reveals the energy flux, $F(u)$, emitted by the mirror \cite{Davies:1976hi},
\be F(u) = -\frac{1}{24\pi}\{p(u),u\}, \label{F(u)}\ee
where $p(u)$ is the advanced time as a function of retarded time, \cite{Davies:1977yv, Birrell:1982ix}. However, since we employ $f(v)$, retarded time as a function of advanced time, Eq.~(\ref{f(v)}), \cite{Good:2016atu,Good:2020byh} we need
\be F(v)= \frac{1}{24\pi}\{f(v),v\}f'(v)^{-2},\label{F(v)}\ee
where the Schwarzian brackets are defined as usual,
\be \{f(v),v\}\equiv \frac{f'''}{f'} - \frac{3}{2}\left(\frac{f''}{f'}\right)^2\,,\ee 
which gives, using $f(v)$ of Eq.~(\ref{f(v)}), 
\be F(v) = \frac{\kappa ^2 (4 \kappa  v (\kappa  v-1) (\kappa  v (\kappa  v-1)+2)+1)}{48 \pi  (1-2 \kappa  v)^4}.\label{F(v)exact}\ee
The leading order terms of the energy flux at both $v\to-\infty$ and $v\to 0^-$ are respectively,
\be F_{\textrm{past}} = \frac{\kappa ^2}{192\pi }, \qquad F_{\textrm{horizon}} = \frac{\kappa ^2}{48 \pi },\ee
demonstrating the asymptotic fluxes are constant (thermal) and differ by a factor of 4.
A plot of the energy flux, $F(v)$, as a function of advance time $v$ is given in Figure \ref{Fig4}. 

\section{Particle Spectrum}\label{sec:particles}
\begin{figure}[H]
\centering
\includegraphics[width=3.1 in]{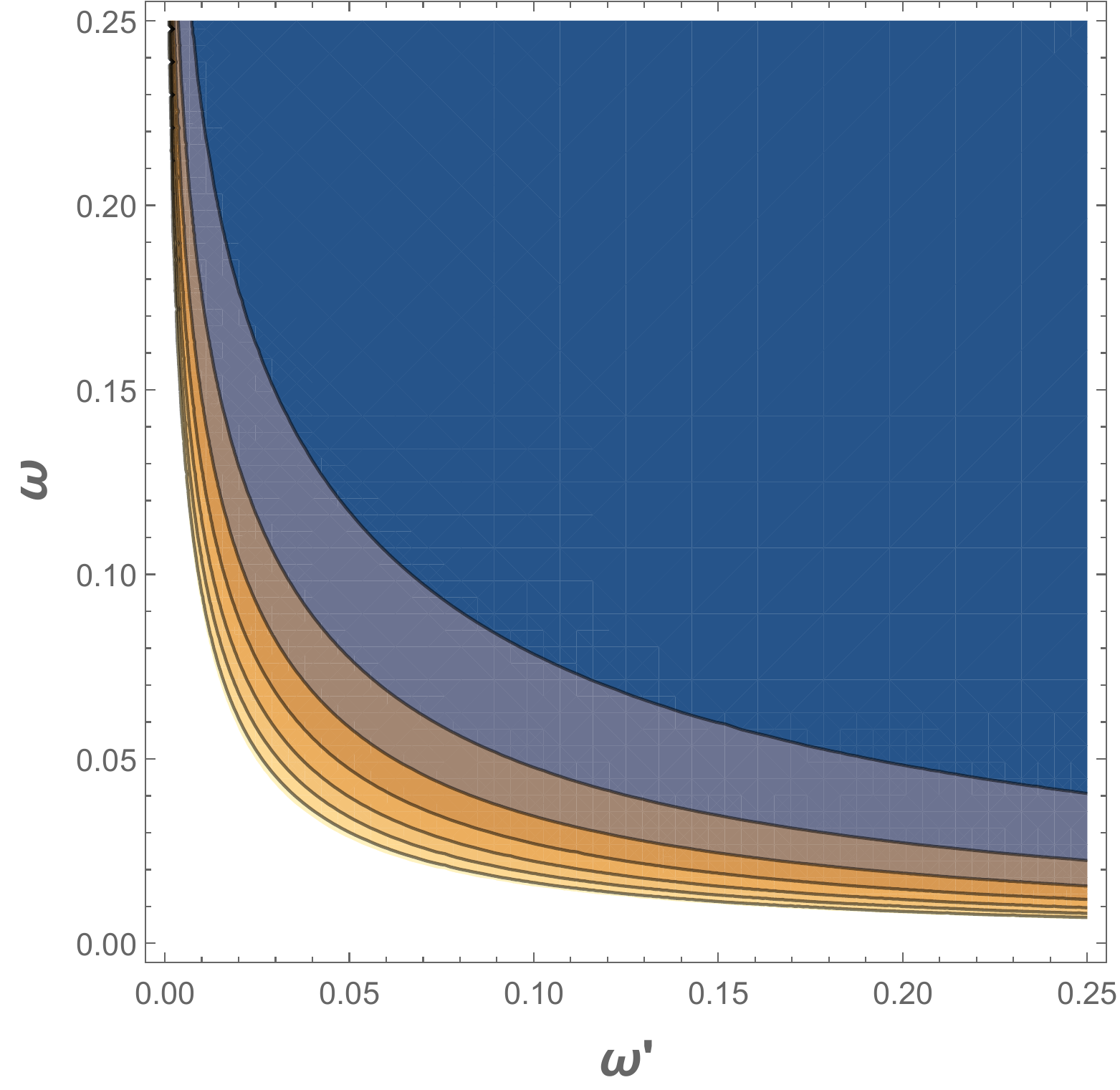}
\caption{The dual-temperature spectrum, Eq.~(\ref{spectrum}), $|\beta_{\omega\omega'}|^2$ as a contour plot, here $\kappa=1$. The color scheme depicts the lightest color as one magnitude order larger than the darkest color. The key take-away is the asymmetry between the two frequencies, $\omega$ and $\omega'$, which underscore the temperature difference between the past-future equilibrium states. }\label{Fig5}
\end{figure}

The particle spectrum can be obtained from the beta Bogoliubov coefficient \cite{Birrell:1982ix},
\be \beta_{\omega\omega'} = \frac{1}{2\pi}\sqrt{\frac{\omega'}{\omega}}\int_{-\infty}^{v_H=0} \d v\: e^{-i\omega'v-i\omega f(v)}\,,\label{partsint}\ee
where $\omega$ and $\omega'$ are the frequencies of the outgoing and incoming modes respectively \cite{carlitz1987reflections}.
To obtain the particle spectrum, we take the modulus square, $N_{\omega \omega'} \equiv |\beta_{\omega\omega'}|^2$, which gives 
\be  N_{\omega \omega'} =\frac{\left|K_{\frac{1}{2}-\frac{i \omega }{\kappa }}\left(\frac{i \omega '}{2 \kappa }\right) \right|^2}{2 \pi ^2 \kappa ^2 \left(e^{2 \pi  \omega/\kappa }-1\right)}.\label{spectrum}\ee
The spectrum Eq.~(\ref{spectrum}), $|\beta_{\omega\omega'}|^2$ is thermal asymptotically and plotted as a contour plot in Figure \ref{Fig5}. In both the early-time ($\omega'\ll \omega$) and late-time ($\omega'\gg \omega$) regimes, we have to leading order, respectively,
\be N_{\omega \omega'}^{\textrm{past}} = \frac{1}{\pi\omega'\kappa}\frac{1}{e^{4 \pi  \omega/\kappa }-1},\quad N_{\omega \omega'}^{\textrm{horizon}}= \frac{1}{2\pi\omega'\kappa} \frac{1}{e^{2 \pi  \omega/\kappa }-1}.\ee
This demonstrates that the particles at early-times (horizonless) have a different temperature than at late-times (horizon). 
This spectrum, Eq.~(\ref{spectrum}), demonstrates a new form of Hawking radiation emanating from a moving mirror trajectory.  Here the temperatures are respectively, 
$T_{\textrm{past}} = \kappa/(4\pi)$, and  $T_{\textrm{horizon}} = \kappa/(2\pi)$.
Eq.~(\ref{spectrum}) can be compared to the late time spectra of non-extremal black holes which is the same expression as $N_{\omega\omega'}^{\textrm{horizon}}$ (e.g. see Eq. 2.13 of \cite{good2013time}) while the Bessel function is reminiscent of the radiation characterizing extremal black holes at lates times \cite{good2020extreme, Liberati:2000sq,Foo:2020bmv}.

The small frequency infrared divergence demonstrates the infinite soft total particle count (some of which are zero-energy Rindler particles \cite{Cozzella:2020gci,Landulfo:2019tqj}) commonly associated with drifting moving mirror solutions \cite{Good:2016atu,Good:2018ell,Good:2018zmx,Myrzakul:2018bhy,Good_2015BirthCry,Good:2016yht} (remnant analogs), that are not strictly asymptotically static  \cite{Walker_1982, Good:2019tnf,GoodMPLA,Good:2017kjr,good2013time,Good:2017ddq,Good:2018aer} (effervescent analogs).  
\section{Conclusions}
We have presented a model of acceleration radiation with an exactly solvable spectrum and energy flux whose particles are asymptotically in a Planck distribution and energy which is a constant equilibrium emission.  The different asymptotic temperatures explicitly demonstrate the disconnect between horizon and non-horizon radiating states of the system.   

Starting with the motion we have used the well-established moving mirror model to compute the precise nature of the radiation, which we liken to a type of `mirror defogger'.  In practise, a store-bought mirror defogger is used to transmit gentle warmth across the mirror's surface, preventing steam's moisture from collecting on the cold mirror.  Our system gently (asymptotically) warms up so that the moving mirror's radiation has an increased temperature, preventing particles from maintaining their original ultra-relativistic cold equilibrium in the asymptotic past.  It moves in a precise non-thermal way, with dynamics that asymptotically evolve to a higher temperature, re-establishing an ultra-relativistic hot equilibrium in the asymptotic future.  The spectral solution describes non-equilibrium emission similar to that in early-time gravitational collapse or the particle creation in cosmologies where the asymptotic early universe and remote future exhibit thermal equilibrium. 

Future studies investigating universal out-of-equilibrium behaviour of Hawking radiation and comparison with other approaches, like holographic ones \cite{Sonner:2012if} may prove fruitful. Regardless, recent proposals involving  phenomenological issues say, e.g. concerning Hawking points in the sky \cite{An:2018utx},
 or cosmological consequences of PBHs make apparent the need for detailed study of vacuum radiation with a variety of possible methods including accelerating mirror models.


\section*{Funding}

MG is funded by grant no. BR05236454 by the Ministry of Education and Science of the Republic of Kazakhstan and the FY2018-SGP-1-STMM Faculty Dev. Competitive Research Grant No. 090118FD5350.
AM acknowledges the support of Grant No. 110119FD4534.

\pagebreak

\clearpage


\end{document}